\documentclass[sigconf]{acmart}
 
\usepackage[english]{babel}
\usepackage{booktabs}
\usepackage{mathrsfs}
\usepackage{multirow}
\usepackage{array,ragged2e}

\usepackage{algorithm}
\usepackage{algpseudocode}
\usepackage{amsmath}
\usepackage{bm}
\usepackage{amssymb}
\usepackage{enumitem}

\usepackage{tikz}
\usetikzlibrary{arrows.meta}
\usetikzlibrary{backgrounds}
\usetikzlibrary{positioning, fit, mindmap, trees, calc,tikzmark,shapes}
\usetikzlibrary{shapes.arrows, fadings, automata,tikzmark,decorations.pathreplacing,patterns}
\usepackage{pgfplots}
\usepgfplotslibrary{groupplots}
\pgfplotsset{compat=1.14}

\usepackage{xcolor}

\definecolor{riptide}{RGB}{141,211,199}
\definecolor{pale_prim}{RGB}{255,255,179}
\definecolor{lavender_gray}{RGB}{190,186,218}
\definecolor{salmon}{RGB}{242,131,107}
\definecolor{seagull}{RGB}{128,177,211}
\definecolor{rajah}{RGB}{253,180,98}
\definecolor{yellow_green}{RGB}{198,222,119}
\definecolor{classic_rose}{RGB}{252,205,229}
\definecolor{feijoa}{RGB}{178,223,138}

\definecolor{cruise}{RGB}{179,226,205}
\definecolor{apricot}{RGB}{253,205,172}
\definecolor{periwinkle}{RGB}{203,213,232}
\definecolor{snow_flurry}{RGB}{230,245,201}
\definecolor{buttermilk}{RGB}{255,242,174}

\definecolor{sundown}{RGB}{249, 180, 181}
\definecolor{spindle}{RGB}{179,205,227}
\definecolor{tea_green}{RGB}{204,235,197}
\definecolor{languid_lavender}{RGB}{222,203,228}
\definecolor{champagne}{RGB}{254,217,166}
\definecolor{cream}{RGB}{255,255,204}

\definecolor{nonte_carlo}{RGB}{135,204,194}
\definecolor{melon}{RGB}{254,191,181}
\definecolor{granny_smith_apple}{RGB}{150,214,150}
\definecolor{mona_lisa}{RGB}{246,152,134}
\definecolor{watusi}{RGB}{254,221,207}
\definecolor{see_green}{RGB}{161,228,195}

\definecolor{moss_green}{RGB}{170,216,176}
\definecolor{opal}{RGB}{164,207,190}

\definecolor{pale_turquoise}{RGB}{172,240,242}
\definecolor{Madang}{RGB}{190,235,159}
\definecolor{pixie_green}{RGB}{183,214,170}
\definecolor{coral_andy}{RGB}{243,204,205}
\definecolor{manhattan}{RGB}{226,180,125}
\definecolor{quartz}{RGB}{219,223,238}
\definecolor{spring_sun}{RGB}{242,243,195}
\definecolor{dairy_cream}{RGB}{254,226,189}
\definecolor{surf_crest}{RGB}{205,230,208}
\definecolor{french_pass}{RGB}{195,232,246}
\definecolor{cosmos}{RGB}{248,209,210}
\definecolor{portafino}{RGB}{245,237,160}
\definecolor{sail}{RGB}{163,205,235}
\definecolor{hint_green}{RGB}{226,246,209}

\definecolor{jet_stream}{RGB}{188, 214, 210}

\definecolor{azalea}{RGB}{251, 196, 196}
\definecolor{wewak}{RGB}{244, 143, 150}
\definecolor{bittersweet}{RGB}{255,111,105}
\definecolor{sunset_orange}{RGB}{242,89,75}
\definecolor{light_coral}{RGB}{244, 127, 123}
\definecolor{carnation}{RGB}{245, 80, 86}
\definecolor{flamingo}{RGB}{237, 88, 85}
\definecolor{carmine_pink}{RGB}{231, 76, 60}
\definecolor{deep_carmine_pink}{RGB}{236, 50, 67}
\definecolor{fire_engine_red}{RGB}{210,44,41}
\definecolor{amaranth}{RGB}{234,46,73}
\definecolor{ku_crimson}{RGB}{243, 0, 25}
\definecolor{fire_engine_red}{RGB}{206, 37, 51}
\definecolor{copper_rust}{RGB}{155, 64, 74}

\definecolor{chilean_fire}{RGB}{215, 87, 44}

\definecolor{japanese_laurel}{RGB}{53, 116, 40}

\definecolor{turmeric}{RGB}{211, 178, 76}
\definecolor{saffron}{RGB}{249,193,62}
\definecolor{my_sin}{RGB}{255, 176, 59}
\definecolor{tree_poppy}{RGB}{246, 154, 27}
\definecolor{jaffa}{RGB}{240, 131, 58}
\definecolor{crusta}{RGB}{254, 127, 44}
\definecolor{tahiti_gold}{RGB}{223, 102, 36}
\definecolor{outrageous_orange}{RGB}{255, 100, 45}
\definecolor{safety_orange}{RGB}{254, 106, 0}

\definecolor{turquoise}{RGB}{41,217,194}
\definecolor{puerto_rico}{RGB}{94, 194, 166}
\definecolor{mountain_meadow}{RGB}{0, 163, 136}
\definecolor{free_speech_aquamarine}{RGB}{0, 156, 114}
\definecolor{java}{RGB}{2,190,196}

\definecolor{matisse}{RGB}{25, 104, 167}
\definecolor{shakespeare}{RGB}{85, 154, 193}
\definecolor{mona_lisa}{RGB}{246,152,134}

\definecolor{bgc}{RGB}{245,245,245}
\definecolor{tuatara}{RGB}{67, 67, 67}
\definecolor{aluminum}{RGB}{153,153,153}
\definecolor{silver}{RGB}{191,191,191}
\definecolor{platinum}{RGB}{228,228,228}
\definecolor{mercury}{RGB}{230,230,230}
\definecolor{gallery}{RGB}{240,240,240}
\definecolor{athens_gray}{RGB}{236, 240, 241}

\definecolor{early_dawn}{RGB}{252,243,218}
\definecolor{egg_shell}{RGB}{238, 234, 215}
\definecolor{midnight}{RGB}{0, 29, 50}
\definecolor{sundown}{RGB}{249, 180, 181}
\definecolor{sun_shade}{RGB}{255, 144, 68}
\definecolor{sushi}{RGB}{117, 168, 47}
\definecolor{tomato}{RGB}{255, 97, 56}
\definecolor{ice_cold}{RGB}{169,232,220}

\definecolor{jelly_bean}{RGB}{45, 126, 150}
\definecolor{shakespeare}{RGB}{85, 154, 193}
\definecolor{celestial_blue}{RGB}{52, 152, 219}
\definecolor{curious_blue}{RGB}{41, 128, 185}
\definecolor{french_blue}{RGB}{0, 112, 182}
\definecolor{matisse}{RGB}{25, 104, 167}

\definecolor{biscay}{RGB}{44, 62, 80}

\definecolor{cosmic_latte}{RGB}{222, 247, 229}
\definecolor{chinook}{RGB}{163, 232, 178}
\definecolor{padua}{RGB}{121, 189, 143}
\definecolor{ocean_green}{RGB}{79, 176, 112}
\definecolor{pastel_green}{RGB}{107, 227, 135}
\definecolor{chateau_green}{RGB}{69, 191, 85}
\definecolor{RoyalBlue}{RGB}{69, 191, 85}
\definecolor{pigment_green}{RGB}{0, 175, 79}
\definecolor{fern}{RGB}{101,197,117}
\definecolor{killarney}{RGB}{56, 113, 66}
\definecolor{viridian}{RGB}{70, 137, 102}

\definecolor{jet_stream}{rgb}{0.69,0.61,0.85}
\definecolor{jelly_bean}{rgb}{0.47,0.32,0.66}

\makeatletter
\g@addto@macro\normalsize{%
	\abovedisplayskip 3pt plus 2pt minus 2pt%
	\belowdisplayskip \abovedisplayskip
	\abovedisplayshortskip 3pt plus2pt  minus2pt%
	\belowdisplayshortskip 3pt plus2pt minus2pt%
}

\makeatother

\usepackage{balance}

\def\BibTeX{{\rm B\kern-.05em{\sc i\kern-.025em b}\kern-.08emT\kern-.1667em\lower.7ex\hbox{E}\kern-.125emX}}

\copyrightyear{2019}
\acmYear{2019}
\setcopyright{acmcopyright}
\acmConference[KDD '19] {The 25th ACM SIGKDD Conference on Knowledge Discovery and Data Mining}{August 4--8, 2019}{Anchorage, AK, USA}
\acmPrice{15.00}
\acmDOI{10.1145/3292500.3330652}
\acmISBN{978-1-4503-6201-6/19/08}

\begin{document}

\copyrightyear{2019} 
\acmYear{2019} 
\setcopyright{acmlicensed}
\acmConference[KDD '19]{The 25th ACM SIGKDD Conference on Knowledge Discovery and Data Mining}{August 4--8, 2019}{Anchorage, AK, USA}
\acmPrice{15.00}
\acmDOI{10.1145/3292500.3330652}
\acmISBN{978-1-4503-6201-6/19/08}

\title{\textit{POG}: Personalized Outfit Generation for Fashion Recommendation at Alibaba \texttt{iFashion}}

\author{Wen Chen, Pipei Huang}
\authornote{Pipei Huang is the Corresponding author.}
\author{Jiaming Xu, Xin Guo, Cheng Guo, Fei Sun, Chao Li, Andreas Pfadler, Huan Zhao, Binqiang Zhao} 

\affiliation{
	\institution{Alibaba Group, Beijing, China}
	{\{chenyu.cw, pipei.hpp, jiaming.xjm, gary.gx, mike.gc, ofey.sf, ruide.lc, andreaswernerrober, chuanfei.zh binqiang.zhao\}@alibaba-inc.com} 
}

\renewcommand{\shortauthors}{Wen Chen and Pipei Huang, et al.}

\begin{abstract}
Increasing demand for fashion recommendation raises a lot of challenges for online shopping platforms and fashion communities.
In particular, there exist two requirements for fashion outfit recommendation: the \textit{Compatibility} of the generated fashion outfits, and the \textit{Personalization} in the recommendation process. 
In this paper, we demonstrate these two requirements can be satisfied via building a bridge between outfit generation and recommendation. 
Through large data analysis, we observe that people have similar tastes in individual items and outfits. 
Therefore, we propose a \textbf{P}ersonalized \textbf{O}utfit \textbf{G}eneration (\textbf{POG}) model, which connects user preferences regarding individual items and outfits with Transformer architecture. 
Extensive offline and online experiments provide strong quantitative evidence that our method outperforms alternative methods regarding both compatibility and personalization metrics. Furthermore, we deploy POG on a platform named \texttt{Dida} in \texttt{Alibaba} to generate personalized outfits for the users of the online application \texttt{iFashion}. 

This work represents a first step towards an industrial-scale fashion outfit generation and recommendation solution, which goes beyond generating outfits based on explicit queries, or merely recommending from existing outfit pools. 
As part of this work, we release a large-scale dataset consisting of 1.01 million outfits with rich context information, and 0.28 billion user click actions from 3.57 million users. 
To the best of our knowledge, this dataset is the largest, publicly available, fashion related dataset, and the first to provide user behaviors relating to both outfits and fashion items.

\end{abstract}



\maketitle

\section{Introduction}

\begin{figure}
	\includegraphics[width=1.0\columnwidth]{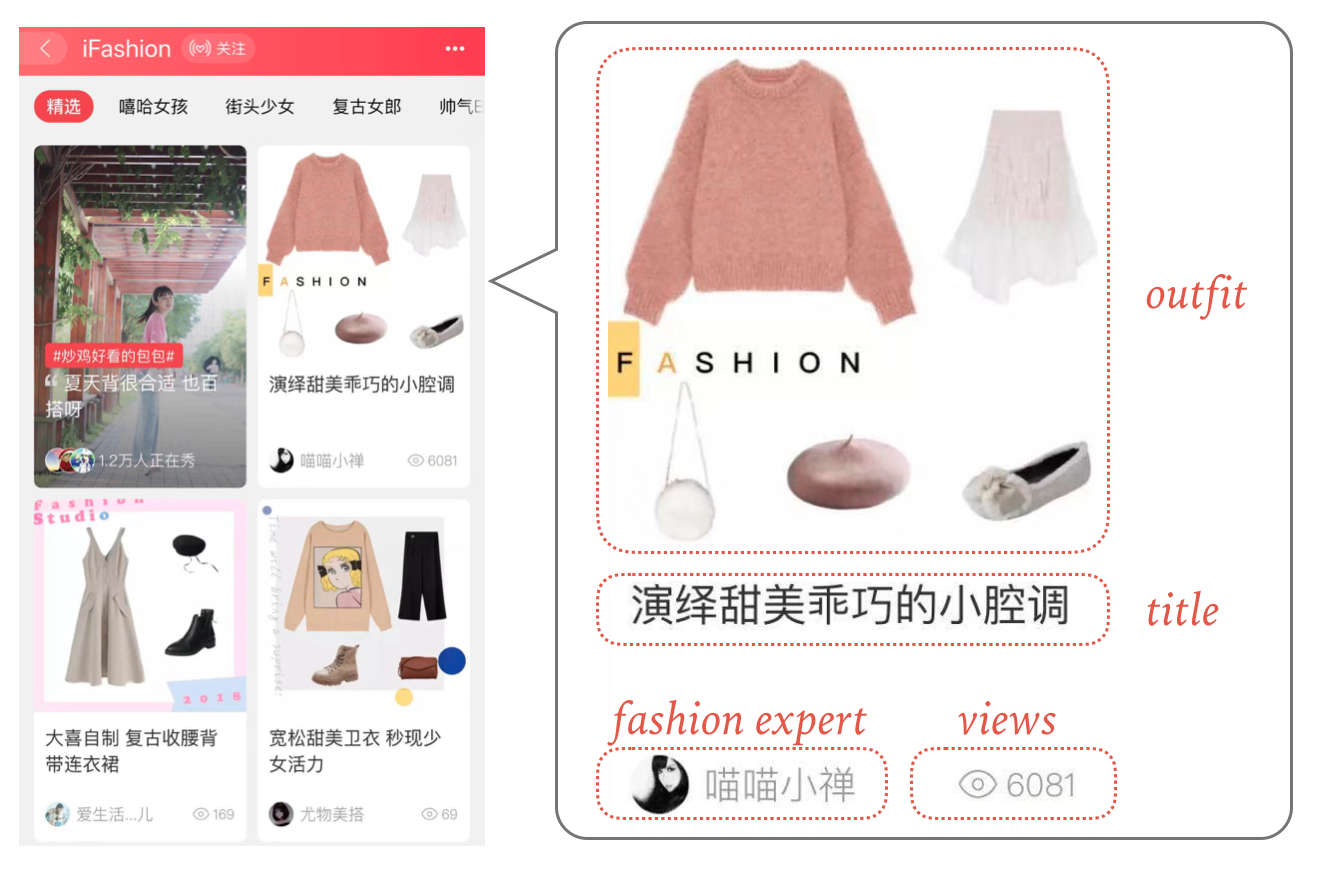}
	\caption{A sample of \texttt{iFashion} application in \texttt{Taobao}. We recommend fashion outfits (sets of fashion items which interact with each other) to users.}
	\label{pic:ifashion}
\end{figure}

Faced with seemingly abundant choices of ever changing styles, fashion outfit recommendation becomes more and more important for modern consumers and has thus attracted interest from the online retail industry. 
Fashion outfit is a set of fashion items, which appears both visually compatible and functionally irredundant \cite{han2017learning} (see Figure \ref{pic:ifashion} as an example). 
Compared to traditional item recommendation, fashion outfit recommendation involves a remarkably creative outfit generation process, which requires both innovation and characteristic.
Therefore, this task is usually performed by fashion experts, and becomes popular in numerous online fashion communities, such as Lookbook\footnote{https://lookbook.nu/} and Chictopia\footnote{http://chictopia.com/}.
In \texttt{Taobao}, the largest online consumer-to-consumer platform in China, a new application \texttt{iFashion} is created to support fashion outfit recommendation.
Approximately 1.5 million content creators were actively supporting \texttt{Taobao} as of March 31, 2018\footnote{https://www.alibabagroup.com/en/news/press\_pdf/p180504.pdf}. 
Still, the great gap between limited human labor and ever-growing market demands a complementary or even substitution of manual work. To alleviate this problem, we aim to assist the generation and recommendation process in \texttt{iFashion}. 

There exist two requirements in fashion outfit generation and recommendation: \textbf{1)} the \textit{Compatibility} of the generated fashion outfits, \textbf{2)} the \textit{Personalization} in the recommendation process. 
\textit{Compatibility} is a measurement of how harmonious a set of items is.
Early studies mainly focus on learning compatibility metric between pairwise items~\cite{veit2015learning,mcauley2015image}, or predicting the popularity of an outfit~\cite{li2017mining}. 
Some recent works attempt to generate outfits by modeling an outfit as an ordered sequence of items~\cite{han2017learning,nakamura2018outfit}. 
However, it is not reasonable because shuffling items in the outfit should make no difference on its compatibility.
\textit{Personalization} represents how the recommendations meet users' personal fashion tastes. 
In recent works, personalization is achieved relying on explicit input (\textit{e.g.}, image or text) provided by the user~\cite{liu2012hi,hu2015collaborative, nakamura2018outfit}. 
This type of fashion generation works more like search than recommendation since it needs explicit user queries.
On the other hand, classic Collaborative Filtering (CF) methods~\cite{sarwar2001item,linden2003amazon} in recommendation mostly focus on recommending individual items rather than outfits.

\begin{table} 
	\caption{The percentage of the same brand, category, style, and pattern of the items in users' clicked outfits appearing in users' latest clicked items.}
	\label{tab:property}
	\begin{tabular}[]{c| c c c c}
		\toprule
		Property & brand & category & style & pattern \\
		\midrule
		Percentage & 56.9\% & 81.3\% & 73.0\% & 53.9\% \\ 
		\bottomrule
	\end{tabular}
\end{table}

In this work, we collect 1.21 billion user clicks on 4.68 million items and 192 thousand outfits from 5.54 million users in \texttt{iFashion}. 
As shown in Table~\ref{tab:property}, 81.3\% of the clicked outfits contain the items with the same categories which have been appeared in the latest 50 clicked items from the same user. 
We observe that the brand, style, and pattern of the items in users' clicked outfits also have high probabilities to appear in users' latest clicked items(we choose the latest 50 items for each user).
Figure~\ref{pic:intro-user-behavior} illustrates three example users with their behaviors on items and outfits. 
All these results show that users tend to keep similar tastes in individual items and outfits. 
We find that although fashion outfit generation and recommendation have been studied intensively in recent years, existing works usually study these two requirements separately. 

Therefore, we attempt to build the bridge between fashion outfit generation and recommendation in a real-world application with millions of users.
More specifically, we generate personalized outfits by capturing users' interests and tastes from their historical interactions on fashion items.
For the \textit{Compatibility} requirement, we propose a \textbf{F}ashion \textbf{O}utfit \textbf{M}odel (FOM) by learning the compatibilities between each item and all the other items within the outfit.
Each item should have different weighted interactions to the other items in the outfit.
Thus, we set up a masked item prediction task based on the self-attention mechanism~\cite{vaswani2017attention}, which masks one item at a time in the outfit, and predicts the masked item based on the context from other items in the outfit.
For the \textit{Personalization} requirement, 
by integrating user preference into the pre-trained FOM, we propose a \textbf{P}ersonalized \textbf{O}utfit \textbf{G}eneration (POG) model, which can generate compatible and personalized outfits based on users' recent behaviors.
Specifically, POG uses a Transformer encoder-decoder architecture~\cite{vaswani2017attention} to model both signals from user preference and outfit compatibility.
To the best of our knowledge, this is the first study to generate personalized outfits based on users' historical behaviors with an encoder-decoder framework.
Last, we develop a platform named \texttt{Dida}, where POG has been deployed, to assist outfit generation and recommendation in large-scale online application \texttt{iFashion}. 

The contributions of this work are summarized as follows:

\begin{figure}
	\includegraphics[width=1.0\columnwidth]{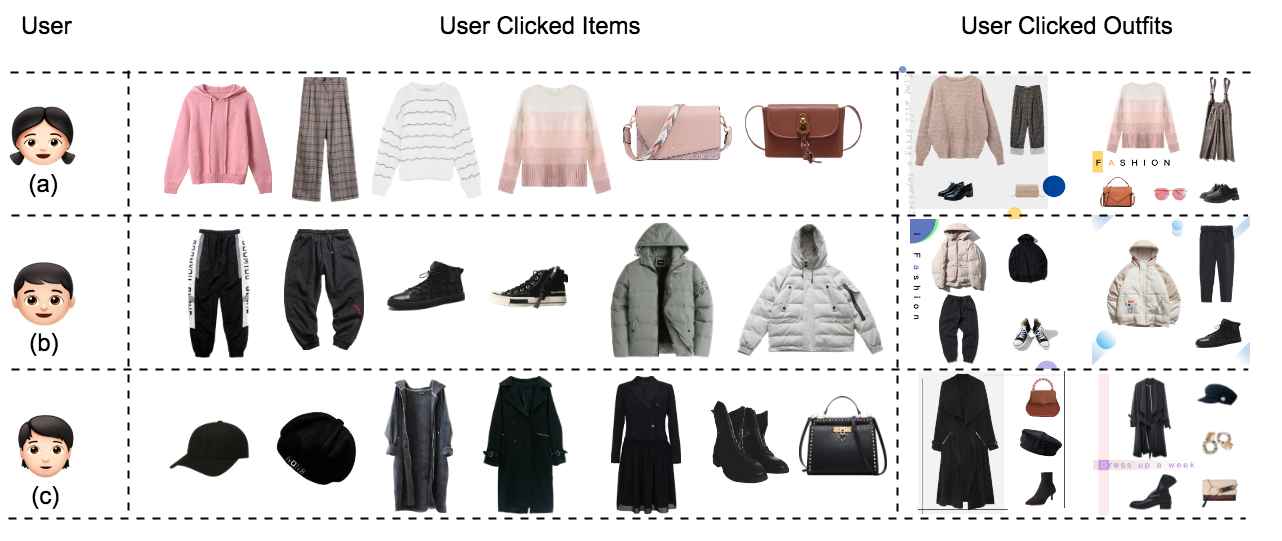}
	\caption{Illustration of user clicked items and outfits. User (a) is a young girl who likes clothes and outfits in light colors and sweet styles. User (b) is a college boy who clicks several winter outfits after clicking a lot of winter clothes. User (c) is probably an office lady who prefers OL style items as well as outfits.}
	\label{pic:intro-user-behavior}
\end{figure}

\begin{enumerate}%
\item We propose POG: an encoder-decoder model to generate personalized fashion outfits, which takes into account both outfit compatibility and user personalization. It makes fashion recommendations by generating personalized outfits based on users' recent behaviors.
\item We demonstrate that our model significantly outperforms other alternative methods through outfit compatibility experiments, including pushing the FITB (Fill In The Blanks) benchmark to 68.79\% (5.98\% relative improvement) and CP (Compatibility Prediction) benchmark to 86.32\% (25.81\% relative improvement).

\item We deploy POG on the real-world outfit generation platform \texttt{Dida}. Through extensive online experiments, we show that POG clearly outperforms the CF method by 70\% increase in CTR (Click-Through-Rate) metric.

\item We release a dataset\footnote{https://github.com/wenyuer/POG} of 1.01 million outfits, 583 thousand fashion items associated with rich context information, and 0.28 billion user click actions from 3.57 million users. 
\end{enumerate}

\section{Related Work}
Fashion is an important application domain of computer vision and multimedia. Much research effort has been made in this domain, focusing on fashion image retrieval \cite{yamaguchi2013paper,hadi2015buy}, clothing recognition \cite{liu2016deepfashion}, clothing parsing \cite{yamaguchi2012parsing,liang2016clothes}, attribute learning \cite{liu2012hi,huang2015cross}, outfit compatibility \cite{han2017learning,veit2015learning,mcauley2015image,vasileva2018learning,tautkute2018deepstyle}, and fashion recommendation \cite{hu2015collaborative,kang2017visually,liu2012hi}. The goal of this work is to compose personalized fashion outfits automatically based on user behaviors, we hence focus on the research areas of outfit generation and recommendation.

\subsection{Fashion Outfit Generation}
Methods for fashion outfit generation usually fall within one of two categories. Methods in the first category focus on calculating a pairwise compatibility metric \cite{veit2015learning,mcauley2015image,song2018neural}.
McAuley et al. extract visual features to model human visual preference for a pair of items of the Amazon co-purchase dataset \cite{mcauley2015image}. 
Siameses network \cite{veit2015learning} estimates pairwise compatibility based on co-occurrence in large-scale user behavior data. 
Methods belonging to the second category, such as presented in \cite{li2017mining} and \cite{han2017learning}, are based on modeling a fashion outfit as a set or an ordered sequence. 
Li et al. deploy an end-to-end deep learning system which can classify a given outfit as popular or unpopular \cite{li2017mining}. 
Han et al. train a bidirectional LSTM model to sequentially generate outfits \cite{han2017learning}. 
These methods generally use simple pooling of item vectors to represent an outfit, or rely heavily on the order of the outfit items. 
We note that methods belonging to either category hardly consider all the interactions among the items in an outfit. 
Besides, it is unreasonable to regard an outfit as an ordered sequence, because shuffling items in the outfit should make no difference on its compatibility. 
We try to explicitly incorporate this into our modeling architecture by requiring that each item should have different interaction weights with respect to other items in one outfit. 
For example, a ``red shirt'' should have a higher interaction weight with ``blue jeans'', but a smaller weight with a pair of ``white gloves''.

\subsection{Fashion Outfit Recommendation}
Early works on recommendation typically use collaborative filtering to model users' preferences based on their behavior histories \cite{sarwar2001item,linden2003amazon}. 
However, previous works mostly restrict attention to recommending individual items. 
There are a few approaches for recommending the whole fashion outfits. 
Si Liu et al. \cite{liu2012hi} propose an occasion-oriented clothing recommendation method based on attributes and categories. 
The work in \cite{mcauley2015image} and \cite{han2017learning} requires image or text queries to find complementary clothes. 
A functional tensor factorization approach is used to suggest sets of items to users in \cite{hu2015collaborative}. 
As these approaches all require user queries or user uploaded data as input, they are likely to be perceived as less user-friendly by a typical user. 
Moreover, we note that these approaches are rather impractical to implement in an efficient manner in a large-scale online recommender system.

\subsection{Self-Attention and Transformer}
Self-attention is an attention mechanism relating different positions of a single sequence \cite{vaswani2017attention}, which has been used successfully in a variety of tasks \cite{parikh2016decomposable, lin2017structured}. 
Transformer \cite{vaswani2017attention}, a transduction model relying on self-attention, has been widely used and greatly improved the performance for language processing tasks~\cite{devlin2018bert,radford2018improving}. 
In recent language representation research, bidirectional Transformer encoder has exhibited the best performance (BERT \cite{devlin2018bert}) when compared to left-to-right Transformer decoder (OpenAI GPT \cite{radford2018improving}) and bidirectional LSTM (ELMo \cite{peters2018deep}). 
Recently, multi-head self-attention is also introduced to model users' behavior sequences for sequential recommendation~\cite{Kang:ICDM2018:SAS,Fei:bert4rec}.
Different from them, in this paper, we adopt the self-attention to model the compatibility in fashion outfit generation.

\section{Dataset}

\texttt{Taobao}'s fashion experts create thousands of outfits everyday. 
All these manually created fashion outfits are reviewed before they are exhibited online. 
About 1.43 million outfits have so far been created and reviewed in this manner. 
We collect the 80 most frequent leaf categories (\textit{e.g.} sweater, coat, t-shirt, boots, and ring) from the items in all these outfits.
All items in the less frequent categories are removed from the outfits. In the rest of the paper, we only consider the items in these 80 leaf categories.
After that, we filter the outfits which contain fewer than 4 items. 
In total, 1.01 million outfits remained, which are composed of 583 thousand individual items.
 
Moreover, we collect clicks on the items and outfits from \texttt{iFashi}-\texttt{on}'s users in recent three months.
We select click behaviors from 3.57 million active users who have viewed more than 40 outfits in total. The clicks on outfits are recorded only when more than 10 item clicks happened before from the same user. 
Then, we build a training sample by pairing an outfit click with the latest 50 item clicks prior to it from the same user. 
Finally, we obtain 19.2 million training samples which consist of 4.46 million items and 127 thousand outfits. 
Every item in our dataset is associated with white background image, title, and leaf category.

To the best of our knowledge, our dataset is the largest publicly available dataset of fashion items with rich information compared to existing datasets, such as WoW \cite{liu2012hi}, Fashion-136K \cite{jagadeesh2014large}, FashionVC \cite{song2017neurostylist}, Maryland Polyvore \cite{han2017learning}, Polyvore Outfits-D, and Polyvore Outfits \cite{vasileva2018learning}. 
Moreover, we are the first to provide outfit data and associated user behavior data, which can be exploited for future fashion recommendation research.
We provide three datasets describing the outfits, the items, and the user behaviors separately. The statistics of the datasets are shown in Table \ref{tab:dataset-outfit}.

\begin{table}
	\caption{Statistics of the datasets.}
	\label{tab:dataset-outfit}
	\begin{tabular}[width=1.0\columnwidth]{l r r r}
		\toprule
		Dataset & \#Outfits & \#Users & \#Items \\
		\midrule
		Outfit data & 1,013,136 & - & 583,464 \\
		Item data & - & - & 4,747,039 \\
		User data & 127,169 & 3,569,112  & 4,463,302  \\
		\bottomrule
	\end{tabular}
\end{table}

\section{Methodology}
In this section, we introduce our methodology in detail. POG is built within a three-step process: we first embed the items. Second, we build FOM, which learns compatibilities of items within an outfit. Last, once training is completed, we use the resulting pre-trained FOM to initialize POG on a Transformer architecture. 

As a first step, we represent all the items using a multi-modal embedding model. Then we introduce FOM and POG in detail. Last, we 
introduce our \texttt{Dida} platform in this section, which helps to ensure the efficiency and quality requirements in large-scale online application of this work.  

\begin{figure*}
	\includegraphics[width=1.7\columnwidth]{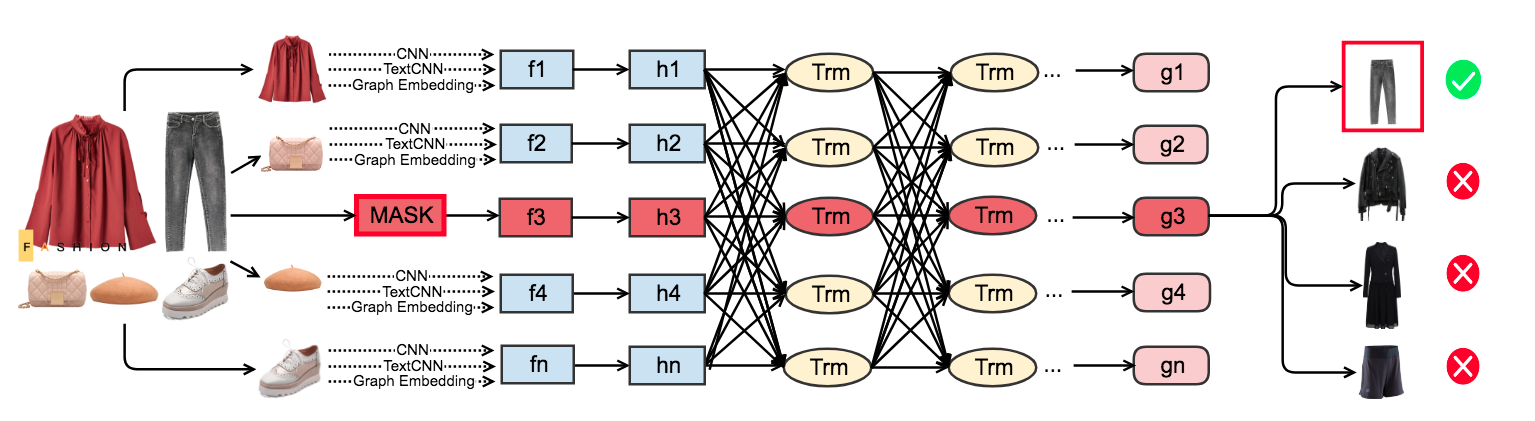}
	\caption{The architecture of FOM. We mask the items in the outfit one at a time. For example, we mask a pair of jeans in the outfit. The model is learned to choose the correct jeans from a candidate pool, to complement other items in the outfit.}
	\label{pic:pretraining}
\end{figure*}

\subsection{Multi-modal Embedding}
\label{sec:embedding}

For every fashion item $f$, we compute a nonlinear feature embedding $\bm{f}$. The concept of fashion relies mostly on visual and textual information. Most previous works suggest to leverage image and text to learn multi-modal embeddings \cite{li2017mining,han2017learning}. In our work, we use a multi-modal embedding model that takes the following input for every item: (1) dense vector encoding the white background picture of the item from a CNN model, (2) dense vector encoding the title of the item obtained from a TextCNN network, which has been pre-trained to predict an item's leaf category based on its title, (3) dense vector encoding a collaborative filtering signal for the item using \texttt{Alibaba}'s proprietary \texttt{Behemoth} Graph Embedding platform \cite{wang2018billion}, which generates item embeddings based on the co-occurrence statistics of items in recorded user click sessions in the \texttt{Taobao} Mobile Application.  

Our goal is to obtain an embedding space, where similar items are embedded nearby, and different items lie in different regions. We concatenate the embeddings derived from image, text, and CF as input for a final fully connected layer. The output is a $d_{e}$-dimension vector $\bm{f}$. For each item $f$, we define positive samples $f^{+}$ as those items which belong to the same leaf category as $f$, and negative samples $f^{-}$ are hence those items which do not fall into the same leaf category. The entire network is then trained using the triplet loss via:

\begin{equation}
\label{eqn:triplet}
\mathcal{L}_{E} = \sum_{\bm{f}} \max \bigl(d(\bm{f},\bm{f^{+}}) - d(\bm{f},\bm{f^{-}}) + \alpha, 0\bigr)
\end{equation}
where the distance metric $d$ represents Euclidean distance, and $\alpha$ as the margin. By minimizing $\mathcal{L}_{E}$, the distance between $f$ and $f^{+}$ in the same category is forced to be smaller than the distance from $f^{-}$ in a different category by some margin $\alpha$. 

\begin{figure*}
	\includegraphics[width=1.6\columnwidth]{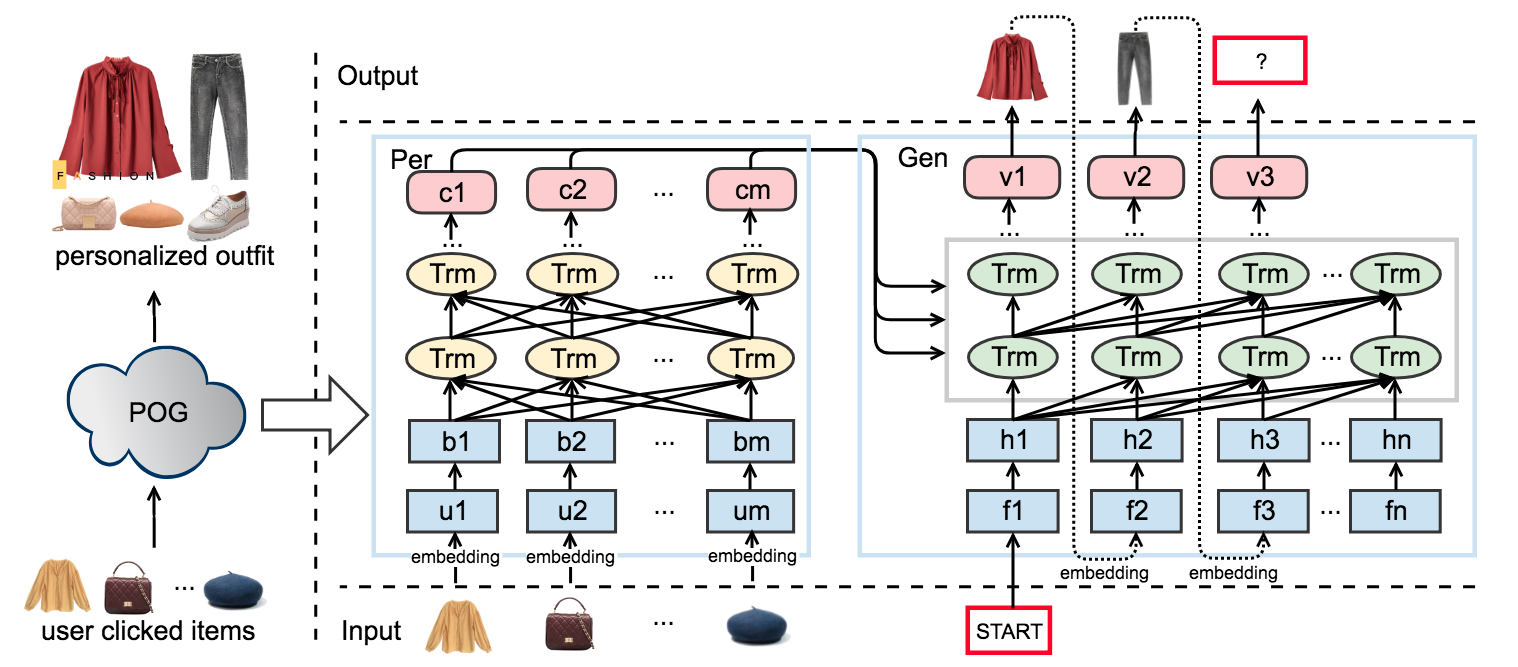}
	\caption{The architecture of POG, which is an encoder-decoder architecture with a \texttt{Per} network and a \texttt{Gen} network. The outfit item is generated step by step according to the user preference signal from the \texttt{Per} network and the compatibility signal from the \texttt{Gen} network.}
	\label{pic:personalized-training}
\end{figure*}

\subsection{FOM: Fashion Outfit Model}
\label{sec:outfit-model}

Outfit is a set of items, where each item should have different weighted interactions to the other items in the outfit. 
To capture the item interactions in an outfit, we design a masked item prediction task based on a bidirectional Transformer encoder architecture.
Masking items one at a time in the outfits, we require the model to fill in the blank with the correct item according to the context. Since every item in the outfit is masked to fuse its left and right context, the compatibility between each item and all the other items within the outfit can be learned from the self-attention mechanism. Hence, the compatibility in the outfit can be learned by combining all theses item compatibilities. 

Let $\mathcal{F}$ denote the set of all outfits. Given an outfit $F \in \mathcal{F}$, $F$ = $\{f_1,\dots,f_t,\dots,f_{n}\}$, where $f_t$ is the $t$-th item. Let $\bm{f}_t$ be the representation of item $f_t$ derived from multi-modal embedding with dimension $d_{e}$. We use a particular embedding [MASK] for the masked item. Non-masked items are represented by their multi-modal embeddings. We then represent the set of input embeddings as $F_{mask}$. Given $F_{mask}$, the task is to predict the masked item rather than reconstructing the entire outfit. More formally, we minimize the following loss function of FOM: 

\begin{equation}
\label{eqn:mask}
\mathcal{L}_F = - \frac{1}{n} \sum_{mask = 1}^{n} \log Pr(f_{mask} | F_{mask}; \bm{\Theta}_{F})
\end{equation}
where $\bm{\Theta}_{F}$ denotes the model parameters, and $Pr(\cdot)$ is the probability of choosing the correct item conditioned on the non-masked items. 

The model architecture is shown in Figure \ref{pic:pretraining}. We do not use position embedding like Transformer \cite{vaswani2017attention} does, because we take the items in the outfit as a set, not a sequence with position information. Let $F_{mask}$ pass through two fully-connected layers with a Rectified Linear Unit (ReLU) activation in between, to transfer all the input embeddings from the single item space to an outfit space. We call the two fully-connect layers as transition layer, and represent the output as $\bm{H}^0$ $\in$ $\mathbb{R}^{n \times d_{m}}$: 

\begin{equation}
 \bm{H}^0 = \texttt{ReLU} (\bigl[\bm{f}_1^{T};\dots;\bm{f}_{n}^{T}\bigl]^{T}\bm{W}_0^{F} +  \bm{b}_0^{F}) \bm{W}_1^{F} + \bm{b}_1^{F} 
\end{equation}
where $\bm{W}_0^{F} \in \mathbb{R}^{d_{e} \times d_{m}}$, $\bm{W}_1^{F} \in \mathbb{R}^{d_{m} \times d_{m}}$, $\bm{b}_0^{F} \in \mathbb{R}^{n \times d_{m}}$, and $\bm{b}_1^{F} \in \mathbb{R}^{n \times d_{m}}$ are learnable parameters. 

The following Transformer encoder contains multiple layers. Each layer contains a Multi-Head self-attention (\texttt{MH}) sub-layer, and a Position-wise Feed-Forward Network (\texttt{PFFN}) sub-layer, where a residual connection is employed around each of the two sub-layers, followed by Layer Normalization (\texttt{LN}). The definitions of \texttt{MH} and \texttt{PFFN} are identical to the paper \cite{vaswani2017attention}. We define the output of the first sub-layer in layer $i$ as $\bm{H}^i_{1}$. Thus each layer $\bm{H}^i$ can be calculated iteratively:

\begin{align}
	&\bm{H}^i = \mathtt{Transformer_e}(\bm{H}^{i-1}), \forall i = 1,\dots,l \\
	&\mathtt{Transformer_e}(\bm{H}^{i-1}) =  \texttt{LN} \bigl( \bm{H}_{1}^{i-1} + \texttt{PFFN}(\bm{H}_{1}^{i-1}) \bigl)  \\
	&\bm{H}_{1}^{i-1} = \texttt{LN} \bigl( \bm{H}^{i-1} + \texttt{MH}(\bm{H}^{i-1},\bm{H}^{i-1},\bm{H}^{i-1}) \bigl)
\end{align}

After $l$ layers, we obtain the output $\bm{G} = \bm{H}^{l}$. Let $\bm{g}_{mask}$ denote the corresponding output of the input [MASK], we then append a softmax layer on top of $\bm{g}_{mask}$ to calculate the probability of the masked item:

\begin{equation}
	Pr(f_{mask} | F_{mask}; \bm{\Theta}_{F}) = \frac {\exp (\bm{g}_{mask} \bm{h}_{mask} ) } {\sum_{\bm{h} \in \mathcal{H}} \exp (\bm{g}_{mask} \bm{h}) }
\end{equation}
where $\bm{h}_{mask}$ is the ground truth transition embedding of the masked item, and $\mathcal{H}$ contains the transition embeddings from all the items in $\mathcal{F}$. One can choose $\mathcal{H}$ to be the whole set of item transition embeddings of all the items, however, this is not practical due to the large number of high-dimensional embeddings. Therefore, we obtain $\bm{h}$ by randomly sampling 3 items from the whole transition set, which are not appeared in the outfit $F$, together with $\bm{h}_{mask}$. This allows the model to learn the compatibility information from the outfit by looking at a diverse set of samples. Note that $\bm{h}_{mask}$ and $\bm{h}$ do not correspond to the original item embeddings, but rather the outputs of the transition layer.

\subsection{POG: Personalized Outfit Generation Model}
\label{sub:POG}

After modeling outfit compatibility, we now consider a generation model which generates personalized and compatible outfit by introducing user preference signals. 
Take the advantage of encoder-decoder structure, we aim to translate an user's historical behaviors to a personalized outfit.
Let $\mathcal{U}$ denote the set of all users and $\mathcal{F}$ be the set of all outfits. We use a sequence of user behaviors $U = \{u_1,\dots,u_i,\dots,u_{m}\}$ to characterize an user, where $u_i$ are the clicked items by the user. $F = \{f_1,\dots,f_t,\dots,f_{n}\}$ is the clicked outfit from the same user, where $f_t$ are the items in the outfit. At each time step, we predict the next outfit item given previous outfit items and user's click sequence on items $U$. Thus for pair $(U,F)$, the objective function of POG can be written as:

\begin{equation}
    \mathcal{L}_{(U,F)} = - \frac{1}{n} \sum_{t = 1}^{n} \log Pr\bigl(f_{t+1} | f_1,\dots,f_t, U; \bm{\Theta}_{(U,F)}\bigr)
\end{equation}
where $\Theta_{(U,F)}$ denotes the model parameters. $Pr(\cdot)$ is the probability of seeing $f_{t+1}$ conditioned on both previous outfit items and user clicked items. 

The model architecture is shown in Figure \ref{pic:personalized-training}. In POG, the encoder takes user clicked items as the input. Given a special token [START], the decoder then generates an outfit by one item at a time. At each step, the model is auto-regressive consuming the previously generated items as input. The generation stops when a special token [END] appears. In the end, an outfit is generated by composing the output items. We name the encoder as \texttt{Per} network, and the decoder as \texttt{Gen} network. To be precise, the \texttt{Per} network provides a user preference signal, and the \texttt{Gen} network can generate outfits based on both personalization signal and compatibility signal. In the \texttt{Per} network, after a transition layer, a Transformer encoder architecture is followed with a stack of $p$ identical layers. Thus the user preference can be obtained via: 

\begin{eqnarray}
 \bm{B}^0 &=& \texttt{ReLU} \Bigl(\bigl[ \bm{u}_1^{T};\dots;\bm{u}_{m}^{T} \bigl]^{T}\bm{W}_0^{U} +  \bm{b}_0^{U}\Bigl) \bm{W}_1^{U} + \bm{b}_1^{U} \\
 \bm{B}^i &=& \mathtt{Transformer_e}(\bm{B}^{i-1}), i = 1,\dots,p
\end{eqnarray} 
where $\bm{B}^0$ is the output of the transition layer. $\bm{W}_0^{U} \in \mathbb{R}^{d_{e} \times d_{m}}$, $\bm{W}_1^{U} \in \mathbb{R}^{d_{m} \times d_{m}}$, $\bm{b}_0^{U} \in \mathbb{R}^{m \times d_{m}}$, and $\bm{b}_1^{U} \in \mathbb{R}^{m \times d_{m}}$ are learnable parameters. The output of the \texttt{Per} network is $\bm{C} = \bm{B}^p$. 

The \texttt{Gen} network is initialized using the aforementioned pre-trained FOM. In the \texttt{Gen} network, after the transition layer, a Transformer decoder architecture is followed. Each Transformer decoder layer contains three sub-layers. The first sub-layer is a Masked Multi-Head self-attention (\texttt{MMH}) mechanism. The masked design ensures that the prediction for a certain item can depend only on the known outputs of previous items. The second sub-layer performs Multi-Head attention (\texttt{MH}) over the output of user behavior $\bm{C}$, which aims at introducing user information to the outfit generation process. The third sub-layer is a Position-wise Feed-Forward Network (\texttt{PFFN}). Similarly, we employ residual connections around each of the sub layers, followed by Layer Normalization (\texttt{LN}). Define the output of the first sub-layer in layer $i$ as $\bm{H}^i_{1}$, and the output of the second sub-layer as $\bm{H}^i_{2}$, then each Transformer decoder layer $\bm{H}^i$ can be calculated iteratively:

\begin{align}
&\bm{H}^0 = \texttt{ReLU} \Bigl([\bm{f}_1^{T};\dots;\bm{f}_{n}^{T}]^{T}\bm{W}_0^{F} +  \bm{b}_0^{F}\Bigr) \bm{W}_1^{F} + \bm{b}_1^{F} \\
&\bm{H}^i = \mathtt{Transformer_d}(\bm{H}^{i-1}), i = 1,\dots,q \\
&\mathtt{Transformer_d}(\bm{H}^{i-1}) = \texttt{LN} \bigl( \bm{H}_{2}^{i-1} + \texttt{PFFN}(\bm{H}_{2}^{i-1}) \bigl) \\
&\bm{H}_{2}^{i-1} = \texttt{LN} \bigl( \bm{H}_{1}^{i-1}+\texttt{MH}(\bm{H}_{1}^{i-1}, \bm{C},\bm{C}) \bigl)  \\
&\bm{H}_{1}^{i-1} = \texttt{LN} \bigl(\bm{H}^{i-1}+\texttt{MMH}(\bm{H}^{i-1},\bm{H}^{i-1},\bm{H}^{i-1}) \bigl) 
\end{align}
where $\bm{H}^0$ is the output of the transition layer, and $q$ is the number of layers of the \texttt{Gen} network. $\bm{W}_0^{F} \in \mathbb{R}^{d_{e} \times d_{m}}$, $\bm{W}_1^{F} \in \mathbb{R}^{d_{m} \times d_{m}}$, $\bm{b}_0^{F} \in \mathbb{R}^{n \times d_{m}}$, and $\bm{b}_1^{F} \in \mathbb{R}^{n \times d_{m}}$ are learnable parameters. The final output of POG is represented as $\bm{V} = \bm{H}^q$. 

In one training sample with one clicked outfit and several clicked items from the same user, the ground truth prediction is the items in the outfit. We define $\bm{h}_{t+1}$ as the ground truth transition vector of the $t$-th output $\bm{v}_t$. Thus the probability of the next item is calculated as:

\begin{equation}
Pr(f_{t+1} | f_1,\dots,f_t, U; \bm{\Theta}_{(U,F)}) = \frac {\exp (\bm{v}_t \bm{h}_{t+1})} {\sum_{\bm{h} \in \mathcal{H}} \exp (\bm{v}_t \bm{h}) }
\end{equation}
Again, $\mathcal{H}$ is composed of 3 randomly selected embeddings from the transition layer, together with the ground truth vector $\bm{h}_{t+1}$. 

In the inference process, for each output vector $\bm{v}_t$, we deploy a similarity search to the candidate pool based on the following objective function:

\begin{equation}
\label{eqn:fitb}
\bm{h}_{t+1} = \arg \max_{\bm{h}_C \in \mathcal{H}_C} \frac {\exp (\bm{v}_t \bm{h}_C)} {\sum_{\bm{h} \in \mathcal{H}_C} \exp (\bm{v}_t \bm{h}) }
\end{equation}
where $C$ is the candidate item set. Then the item with the transition embedding $\bm{h}_{t+1}$ is chosen to be the next item. The generation stops when the [END] mark appears. A detailed generation process is described in Algorithm \ref{alg:generation}.

\begin{algorithm}[t]
	\caption{Personalized Outfit Generation.}
	\label{alg:generation}
	\begin{algorithmic}[1]
		\Require
		Candidate set $C$;
		User clicked item sequence $U$.
		\Ensure
		Personalized fashion outfit $F$.
		\State Transform all items in $C$ into outfit space $\mathcal{H}_C$.
		\State Input $U$ in the \texttt{Per} network.
		\State Input [START] mark in the \texttt{Gen} network.  
		\Repeat
			\State Get the output $\bm{v}_t$ of the $t$-th step.
			\State Output the next item $f_{t+1}$ in $\mathcal{H}_C$ according to Equation \ref{eqn:fitb}.
			\State Input $f_{t+1}$ to the \texttt{Gen} network.
		\Until {(The [END] mark appears.)}
	\end{algorithmic}
\end{algorithm}
	
\subsection{Dida Platform}
\label{sec:dida}
\begin{figure}
	\includegraphics[width=1.0\columnwidth]{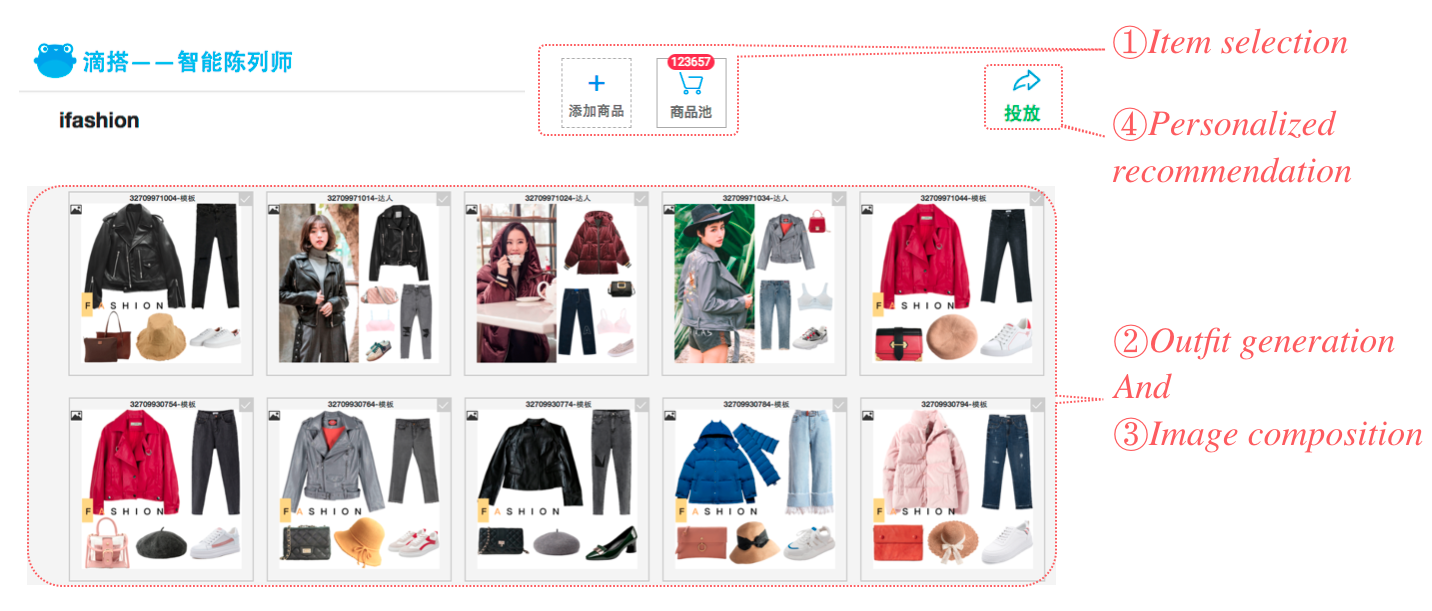}
	\caption{Dida platform. We provide services including item selection, outfit generation, image composition, and personalized recommendation.}
	\label{pic:dida}
\end{figure}

In order to ensure an overall smooth user experience, strict quality and efficiency requirements have to be taken into consideration when deploying POG into an online application with millions of users. 
We thus develop a platform named \texttt{Dida} which is able to generate personalized outfits automatically at very large scales and ensures necessary quality and efficiency standards. 
\texttt{Dida} platform consists of a number of services including item selection, outfit generation, image composition, and personalized recommendation, as shown in Figure \ref{pic:dida}. The outfit generation and recommendation workflow, wherein POG is implemented and deployed, can be described as follows:

\begin{itemize} [labelsep = .5em, leftmargin = 0.1pt, itemindent = 2em]

	\item We support million scale item pools and assist operators to select qualified images with clean background and chosen categories. 
	Images, texts, and CF-data are extracted for selected items and multi-modal embeddings are computed as described in Section \ref{sec:embedding}. 
	In the generation process as described in Algorithm \ref{alg:generation}, we use Faiss \cite{JDH17} to implement the similarity search. 
	To ensure the quality of the generated outfits, we restrict the similarity search according to certain domain rules provided by fashion experts. 
	For example, if it specifies that one t-shirt, one pair of jeans, and one sport shoes as a category rule, then only jeans and sport shoes can be searched in Faiss after the t-shirt. 

	\item The generated items from the first step will be composed together with their images in designed templates. After image composition, we recommend outfits to users. The personalized generated outfits from POG are recommended directly to the users. In addition, we also support other recommendation strategies including random recommendation and CF recommendation.
\end{itemize}

\texttt{Dida} is widely used by more than one million operators at \texttt{Alibaba}. About 6 million personalized outfits are generated everyday with high quality. So far, the outfits have been recommended to more than 5.4 million users.

\section{Experiment}
We describe our experiments and analyze the results in this section. 
We conduct both offline and online experiments to compare the compatibility and recommendation performances.

\subsection{Fashion Outfit Compatibility}

\subsubsection{Implementation Details}
For multi-modal embedding process, we use 1536 dimensional CNN features derived from Inception Resnet V2 model \cite{szegedy2017inception} as the image features. 
The text representations are 300 dimensional vectors derived from TextCNN \cite{kim2014convolutional} with a vocabulary size of 420,758. 
Graph embeddings are 160 dimensional CF vectors from \cite{wang2018billion}. 
We set $d_e = 128$ as the final dimension of the multi-modal embedding. 
Triplet margin is fixed as $\alpha=0.1$. 
In FOM, the model is composed of a stack of $l=6$ layers. 
We use 8 heads in all multi-heads attentions, and $d_m = 64$ as the dimension of hidden layers.

\subsubsection{Task Settings \& Evaluation Metrics}
To evaluate the performances of multi-modal embeddings and models on predicting the outfit compatibility, we adopted two wide-used tasks following the practices in~\cite{han2017learning,li2017mining,nakamura2018outfit,vasileva2018learning}. 

\textbf{Fill In the Blank (FITB)}. FITB is a task predicting an item from multiple choices that is compatible with other items to fill in the blank. 
For offline evaluation, we split 10\% of the data as a test set.
We mask one item at a time for each outfit in the test set. 
Then for each masked item, we randomly select 3 items from other outfits along with the ground truth item to obtain a multiple choice set. 
It is reasonable to believe that the ground truth item is more compatible with other items than the randomly selected ones. 
Finally, the result is evaluated by the accuracy of choosing the correct items.

\textbf{Compatibility Prediction (CP)}.
The CP task is to predict wheth-er a candidate outfit is compatible or not.
For evaluation, we first build a compatible outfit set which is 10\% split from the dataset. 
Then, we produce the same amount of incompatible outfits as the compatible outfits for a balanced test set by randomly selecting fashion items from the compatible outfit set.
In the CP task, a candidate outfit is scored as to whether its constitute items are compatible with each other. 
The performance is evaluated using the AUC (Area Under Curve) of the ROC (Receiver Operating Characteristic) curve. 

Specifically, in FOM, we solve the FITB task based on Equation \ref{eqn:fitb}, where $C$ is the choice set. For CP task, given an outfit, we simply utilize the negative value of Equation \ref{eqn:mask} as the compatibility score.

\subsubsection{Compare Different Modalities}

We compare each modality and their combinations on compatibility evaluations. The performances are compared in Table \ref{tab:multimodal}, which shows that: (1) In both FITB and CP tasks, text works best on its own, but image and CF provide complementary information. (2) CF embedding alone does not work very well, partly because it lacks semantic visual and textual information, which is important in fashion compatibility. (3) The 1536 dimensional CNN feature derived from Inception Resnet V2 is compressed to 128 after the fully connected layer. The resulting dimension is relatively small to contain the important visual information of the fashion items, which partly explains that the image only modality performs badly. Similar results can also been observed from \cite{li2017mining}.

\begin{table}
	\caption{FITB and CP results of different modalities.}
	\label{tab:multimodal}
	\begin{tabular}[width=1.0\columnwidth]{p{2cm}  p{1.5cm}<{\Centering}  p{1.5cm}<{\Centering} }
		\toprule
		Modality & FITB & CP \\
		\midrule
		Image & 62.85\% & 80.78\%\\
		Text & 68.02\% & 85.17\% \\ 
		CF & 51.38\% & 57.92\%   \\  
		Image+text & 68.24\% & 85.41\% \\
		Image+text+CF & \textbf{68.71\%} & \textbf{86.09\%} \\
		\bottomrule
	\end{tabular}
\end{table}

\subsubsection{Compare Different Models}
To demonstrate the effectiveness of our model FOM, we compared several alternative models. For fair comparison, all the models share the same embeddings and the dimension of hidden layers.

\begin{itemize} [labelsep = .5em, leftmargin = 0.1pt, itemindent = 2em]

\item \textbf{F-LSTM}\cite{han2017learning} Given a sequence of fashion items in an outfit, a forward LSTM is trained taking each item as an individual time step. 

\item \textbf{Bi-LSTM}\cite{han2017learning} A bidirectional LSTM adds a backward LSTM compared to F-LSTM, thus predicting the next item can be performed in the reverse order also. 

\item \textbf{SetNN}\cite{li2017mining} SetNN uses a multi-instance pooling model to aggregate information from individual fashion items to produce a holistic embedding for the outfit. We use mean reduction as the pooling method, which performs the best in experiments as \cite{li2017mining} shows. The original SetNN predicts popularity of an outfit by labeling the outfit preferences. Since popularity does not always indicate compatibility, for fair comparison, we change its training inputs with compatible outfits, and incompatible outfits by randomly selecting items from the training set.

\item \textbf{FOM (ours)} Our Fashion Outfit Model by learning a masked item prediction task described in Section \ref{sec:outfit-model}.

\end{itemize} 

In particular, F-LSTM and Bi-LSTM are sequence based models, which take sequences of items as inputs. 
SetNN and FOM are set based models, which do not enforce a specific order over the fashion items as inputs.  
Considering the particular requirement of sequence based models which need ordered inputs, we conduct the experiments with both disordered inputs and ordered inputs. The order follows the guide in \cite{han2017learning}, which has a fixed order: tops, bottoms, shoes, and accessories, where we use leaf categories to specify.

\begin{table}
	\caption{FITB and CP results of different models.}
	\label{tab:fitb}
	\begin{tabular}[width=1.0\columnwidth]{l cc cc}
		\toprule
		\multirow{2}*{Model} & \multicolumn{2}{c}{FITB} & \multicolumn{2}{c}{CP} \\
		\cmidrule(lr){2-3} \cmidrule(lr){4-5}
		& Unordered & Ordered & Unordered & Ordered \\
		\midrule
		F-LSTM\cite{han2017learning} & 58.07\% & 62.84\% & 63.78\% & 65.04\% \\
		Bi-LSTM\cite{han2017learning} & 58.21\% & 64.91\% & 63.82\% & 68.61\%\\
		\midrule
		SetNN\cite{li2017mining} & 49.24\% & 49.27\% & 58.31\% & 58.33\% \\
		\midrule
		FOM (ours) & \textbf{68.71\%} & \textbf{68.79\%} & \textbf{86.09\%} & \textbf{86.32\%} \\
		\bottomrule
	\end{tabular}
\end{table}

\textbf{FITB Results}.
The middle two columns of Table \ref{tab:fitb} shows the results of our model compared with alternative models in the FITB task. From this table, we make the following observations: (1) Sequence based models are sensitive to the input order, while set based models are not. Both F-LSTM and Bi-LSTM have better performances with ordered inputs. SetNN and FOM have similar results with different inputs. This demonstrates that sequence based models can only work well in specific order. (2) Bi-LSTM has better results than F-LSTM both in ordered inputs and unordered inputs. The combination of LSTMs in two directions offers higher accuracy than one directional LSTM. Similar result is also observed in \cite{han2017learning}. (3) FOM performs the best for both inputs. It exhibits a 18.04\% increase compared to the second best result (Bi-LSTM) in unordered inputs, and a 5.98\% increase compared to the second best result (Bi-LSTM) in ordered inputs. We attribute this to the self-attention mechanism in Transformer, which facilitates FOM to calculate the weighted interactions of every item to the others, makes it suitable for modeling compatibility in this task.

\textbf{CP Results}.
The last two columns in Table \ref{tab:fitb} shows the results of our model and others for the CP task. Similar to FITB, sequence based models are still sensitive to the input order. FOM obtains the best performance by significant margins. In unordered inputs, FOM has a 34.90\% increase compared to the second best (Bi-LSTM), while in ordered inputs, FOM shows 25.81\% increase compared to the second best (Bi-LSTM). Although SetNN is directly trained to predict set compatibility, it still performs the worst. Similar results can also be found in \cite{li2017mining}. We attribute the performance of FOM to the multi-head self-attention mechanism. It learns the compatibilities between each item and all the other items within the outfit, and then an outfit compatibility can be learned by combining all these item compatibilities. 

\subsection{Fashion Outfit Generation and Recommendation}
\label{sec:online}
In this section, we conduct extensive online experiments on different fashion models and recommendation methods. The performance of outfit compatibility and recommendation can be evaluated via CTR, which is an explicit metric representing the effectiveness of different models and methods. Further, we show some online example cases based on our model.
\subsubsection{Implementation Details \& Task Settings} 
In POG, the \texttt{Per} network is composed of a stack of $p=6$ layers, and the \texttt{Gen} network is composed of a stack of $q=6$ layers.
We use 8 heads in all multi-head attentions, and $d_m = 64$ as the dimension of hidden layers.
Our online test data is a manually selected item pool with 1.57 million items. Outfits are generated and recommended via different generation models and recommendation methods on \texttt{Dida} platform. We split the users equally into 8 buckets, each tests a certain generation model with recommendation method.

\subsubsection{Compare Different Models and Methods}

Since SetNN \cite{li2017mining} cannot generate outfits, we focus on the comparisons among generation available models. It is worth mentioned that none of the existing generation models, including F-LSTM and Bi-LSTM, are available to do personalized recommendation. So we Randomly Recommend (RR) these generated outfits to the users. To compare the recommendation performance better, we also deploy classic CF \cite{sarwar2001item} methods to recommend. We first estimate a user's most preferred item via measuring its similarities with the items in his or her interaction history by calculating an item-to-item similarity matrix, according to \cite{sarwar2001item}. Then, the model generated outfit which contains the user's most preferred item is recommended. Online experiments are listed as follows:

\begin{itemize} [noitemsep,topsep=0pt, labelsep = .5em, leftmargin = 0.1pt, itemindent = 2em]
	
\item \textbf{F-LSTM\cite{han2017learning}+RR} Outfits are generated by starting with every item in the test set using the F-LSTM model. The generated outfits are recommended randomly to users.

\item \textbf{Bi-LSTM\cite{han2017learning}+RR} Outfits are generated as paper \cite{han2017learning} presents by taking every item in the test set as the query item. The generated outfits are recommended randomly to users.

\item \textbf{\texttt{Gen}+RR (ours)} Only \texttt{Gen} network is used for generation. We input no user information to POG to evaluate the generation quality only. The generated outfits are recommended randomly. 

\item \textbf{F-LSTM\cite{han2017learning}+CF} Outfits are generated by F-LSTM model. The generated outfits are recommended using the CF method.

\item \textbf{Bi-LSTM\cite{han2017learning}+CF} Outfits are generated by Bi-LSTM model. The generated outfits are recommended using the CF method.

\item \textbf{\texttt{Gen}+CF (ours)} Only \texttt{Gen} network is used for generation. The generated outfits are recommended using the CF method.

\item \textbf{POG (ours)} Both the \texttt{Per} network and the \texttt{Gen} network are used in POG to generate and recommend personalized outfits.

\item \textbf{POG+FOM (ours)} The \texttt{Gen} network is initialized using the pre-trained FOM before generating personalized outfits.

\end{itemize}

We record the CTR on outfits over a period of seven days, and report the results in Figure \ref{pic:online}. Three conclusions can be observed from the figure: (1) Although the CTR values vary during the seven days, we still observe an obvious phenomenon that the recommendation performance largely relies on the recommendation methods. CF methods have higher CTR than all the RR methods, and POG methods outperform all the other. POG with pre-trained FOM has the highest CTR in most of the time. It has an improvement of more than 70\% in CTR, compared to the state-of-the-art model Bi-LSTM with CF recommendation. We attribute this to the differences between individual item recommendation and outfit recommendation. The most preferred item of one user may not indicate the most preferred outfit. (2) The compatibility performance of these models matters, but not in an obvious way. This is partly because of the rule based search in the generation process described in Section \ref{sec:dida}, which avoids some extreme bad cases. In most of the time, \texttt{Gen} has the highest CTR both in RR and CF methods. (3) The performance of POG with pre-trained FOM has a little improvement over POG alone. We may conclude that the \texttt{Gen} network in POG is already enough to learn the compatibility of outfits.

\pgfplotsset{
axis background/.style={fill=gallery},
grid=both,
  xtick pos=left,
  ytick pos=left,
  tick style={
    major grid style={style=white,line width=1pt},
    minor grid style=bgc,
    draw=none
    },
  minor tick num=1,
  ymajorgrids,
	major grid style={draw=white},
	y axis line style={opacity=0},
	tickwidth=0pt,
}

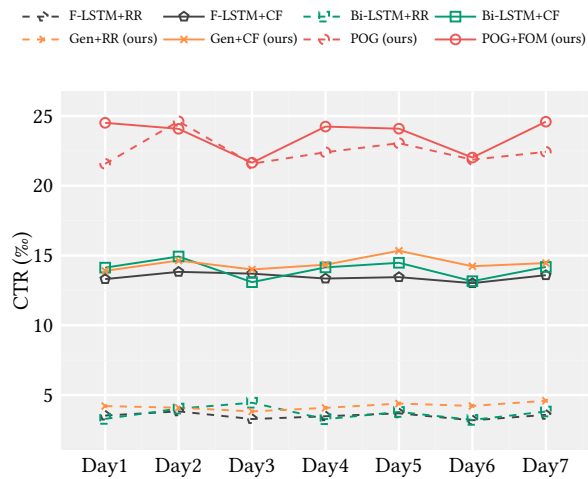
\begin{figure}[t]
\centering
\resizebox{0.95\linewidth}{!}{
    \begin{tikzpicture}[]
	\begin{groupplot}[
    	width=1.09\linewidth,
    	height=0.8\linewidth,
	    group style={group size=1 by 1}, 
        ylabel=CTR (\textperthousand),
        xticklabels={Day1, Day2, Day3, Day4, Day5, Day6, Day7},
        xtick={1,2,3,4,5,6,7},
        ytick={5, 10, 15, 20, 25},
        ymajorgrids,
        major grid style={draw=white},
        y axis line style={opacity=0},
        y label style={at={(-0.04,0.5)}},
        tickwidth=0pt,
        legend cell align={left},
	    ] 
	    \nextgroupplot[
		legend style = {
		  font=\scriptsize ,
          draw=none, 
          fill=none,
          column sep = 0pt, 
          /tikz/every even column/.append style={column sep=0mm},
          legend columns = 4, 
          legend to name = grouplegend},
		]	
		\addplot[thick, dashed, color=tuatara,mark=pentagon] coordinates {
          (1, 3.53)
          (2, 3.83)
          (3, 3.280)
          (4, 3.480)
          (5, 3.690)
          (6, 3.190)
          (7, 3.580)
        }; \addlegendentry{F-LSTM+RR}	
         \addplot[thick, color=tuatara,mark=pentagon] coordinates {
          (1, 13.310)
          (2, 13.830)
          (3, 13.700)
          (4, 13.350)
          (5, 13.450)
          (6, 13.020)
          (7, 13.590)
        }; \addlegendentry{F-LSTM+CF}		
        \addplot[thick, dashed, color=free_speech_aquamarine,mark=square] coordinates {
          (1, 3.290)
          (2, 4.030)
          (3, 4.450)
          (4, 3.270)
          (5, 3.790)
          (6, 3.220)
          (7, 3.820)
        }; \addlegendentry{Bi-LSTM+RR}
        \addplot[thick, color=free_speech_aquamarine,mark=square] coordinates {
          (1, 14.130)
          (2, 14.930)
          (3, 13.090)
          (4, 14.150)
          (5, 14.480)
          (6, 13.160)
          (7, 14.180)
        }; \addlegendentry{Bi-LSTM+CF}		
        \addplot[thick, dashed, color=sun_shade,mark=x] coordinates {
          (1, 4.210)
          (2, 4.090)
          (3, 3.830)
          (4, 4.080)
          (5, 4.390)
          (6, 4.220)
          (7, 4.590)
        }; \addlegendentry{Gen+RR (ours)}	
        \addplot[thick, color=sun_shade, mark=x] coordinates {
          (1, 13.900)
          (2, 14.640)
          (3, 14.000)
          (4, 14.340)
          (5, 15.340)
          (6, 14.230)
          (7, 14.460)
        }; \addlegendentry{Gen+CF (ours)}	
        \addplot[thick, dashed, color=flamingo,mark=o] coordinates {
          (1,21.570)
          (2, 24.620)
          (3, 21.580)
          (4, 22.400)
          (5, 23.050)
          (6, 21.870)
          (7, 22.430)
        }; \addlegendentry{POG (ours)}	
        \addplot[thick, color=flamingo,mark=o] coordinates {
          (1, 24.510)
          (2, 24.080)
          (3, 21.650)
          (4, 24.240)
          (5, 24.090)
          (6, 22.020)
          (7, 24.590)
        }; \addlegendentry{POG+FOM (ours)}
        
    	\end{groupplot}
\node at ($(group c1r1) + (-8pt, 100pt)$) {\ref{grouplegend}};
\end{tikzpicture}}
    \caption{CTR of online experiments. Methods based on POG achieve the best performances. CF based methods are better than RR methods.}
    \label{pic:online}
\end{figure}

\subsubsection{Case Study}

\begin{figure}
	\includegraphics[width=0.9\columnwidth]{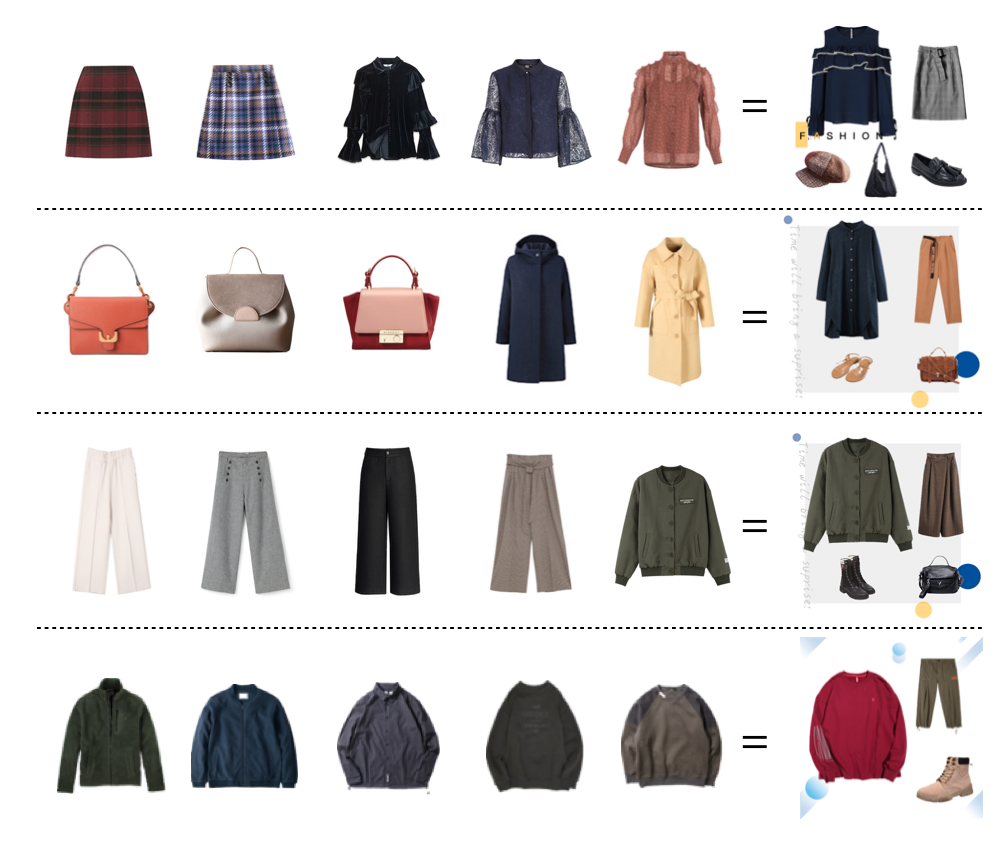}
	\caption{The online cases of POG. Four rows correspond to four users. The first five columns are user clicked items sampled from the latest 50 clicks, and the last column is the generated outfit by POG. }
	\label{pic:case}
\end{figure}

We randomly sample some online cases, and present them in Figure \ref{pic:case}. For users with various preferences on the items, (\textit{e.g.} the first two rows), POG can generate compatible outfits conditioned on the common properties of these items. For users who clicked a lot of similar items, (\textit{e.g.} the last two rows), POG provides an outfit with the main item, which has similar properties with the clicked items. The outfits may inspire the users with better ideas of how to compose the items more properly, and even influence them to buy the whole outfit. 

\section{Conclusion and Future Work}
In this paper, we propose a personalized outfit generation model to build the bridge between compatibility and personalization, which are two essential requirements in fashion generation and recommendation. 
Our model is built in three steps: multi-modal embedding, fashion outfit modeling, and personalized fashion outfit modeling. 
We use Transformer in our model which helps to find interactions of the items in the outfit, and also helps to connect user behaviors on items and outfits. 
Our model outperforms other alternative models in outfit compatibility, outfit generation and recommendation by significant margins. 
Furthermore, we deploy POG on the \texttt{Dida} platform to assist personalized outfit generation in large-scale online application, which is widely used in \texttt{Alibaba} now. We share our data in this paper for future research. 

For the future work, we aim to solve the cold-start problem via building user profiles from similar user groups.
\section*{Acknowledgement}
We would like to thank colleagues of our team - Yuchi Xu, Zhiyuan Liu, Wei Li, Jizhe Wang, Mengmeng Wu, Lifeng Wang, and Qiwei Chen for useful discussion and supports. We are grateful to our cooperative team - Wentao Yang, Jun Bao, Yue Hu, Yumeng Liu, Lijie Cao, Yunfeng Zhang, Qichuan Xiao, and Mingkai Wang. We also thank the reviewers for their valuable comments and suggestions that help improve the quality of this manuscript.



\end{document}